\documentclass[lettersize,journal]{IEEEtran}

\usepackage{amsmath,amsfonts}
\usepackage{algorithmic}
\usepackage[export]{adjustbox}
\usepackage{array}
\usepackage[caption=false,font=normalsize,labelfont=sf,textfont=sf]{subfig}
\usepackage{textcomp}
\usepackage{float}
\usepackage{stfloats}
\usepackage{pifont}
\usepackage{url}
\usepackage{verbatim}
\usepackage{graphicx} 
\usepackage{makecell}
\usepackage{multirow}
\usepackage{supertabular}
\usepackage[singlelinecheck=false]{caption}
\usepackage{listings}
\usepackage[dvipsnames]{xcolor}
\usepackage{url}
\usepackage{soul}
\usepackage{graphicx}
\usepackage{booktabs}
\usepackage{amssymb}
\usepackage{cite}
\usepackage{hyperref}
\usepackage{academicons}
\usepackage{dcolumn}
\usepackage{balance}
\usepackage{setspace}
\usepackage{cite}
\usepackage{algorithm}
\usepackage{algorithmic}
\usepackage{bbm}
\usepackage{bm}
\usepackage{enumitem}
\newcommand{\orcid}[1]{\href{https://orcid.org/#1}{\textcolor[HTML]{A6CE39}{\aiOrcid}}}

\colorlet{punct}{red!60!black}
\definecolor{background}{HTML}{EEEEEE}
\definecolor{delim}{RGB}{20,105,176}
\colorlet{numb}{magenta!60!black}
\def\BibTeX{{\rm B\kern-.05em{\sc i\kern-.025em b}\kern-.08em
    T\kern-.1667em\lower.7ex\hbox{E}\kern-.125emX}}
\usepackage{balance}

\lstdefinelanguage{json}{
    basicstyle=\scriptsize,
    stringstyle=\ttfamily,
    numbers=none,
    showstringspaces=false,
    breaklines=true,
    frame=lines,
    backgroundcolor=\color{background},
}

\newcolumntype{d}[1]{D{.}{.}{#1}}
\newcommand{\rowgroup}[1]{\hspace{-1em}#1}

\colorlet{punct}{red!60!black}
\definecolor{background}{HTML}{EEEEEE}
\definecolor{delim}{RGB}{20,105,176}
\colorlet{numb}{magenta!60!black}
\def\BibTeX{{\rm B\kern-.05em{\sc i\kern-.025em b}\kern-.08em
    T\kern-.1667em\lower.7ex\hbox{E}\kern-.125emX}}

\DeclareMathSymbol{\shortminus}{\mathbin}{AMSa}{"39}

\begin{document}
\include{cite}
\title{Copycat vs. Original: Multi-modal Pretraining and Variable Importance in Box-office Prediction}

\author{\IEEEauthorblockN{Qin Chao\textsuperscript{1,3}, Eunsoo Kim\textsuperscript{2}, and Boyang Li\textsuperscript{1}}\\
\IEEEauthorblockA{chao0009@ntu.edu.sg, eunsoo@uos.ac.kr, boyang.li@ntu.edu.sg} \\
\IEEEauthorblockA{\textit{\textsuperscript{1}College of Computing and Data Science, Nanyang Technological University, Singapore} \\
\textit{\textsuperscript{2}Business School, University of Seoul, Republic of Korea} \\
\textit{\textsuperscript{3}Alibaba Group and the Alibaba-NTU Joint Research Institute, Singapore}
}
}

\markboth{IEEE TRANSACTIONS ON MULTIMEDIA}%
{Copycat vs. Original: Multi-modal Pretraining and Variable Importance in Box-office Prediction}

\maketitle

\begin{abstract}
The movie industry is associated with an elevated level of risk, which necessitates the use of automated tools to predict box-office revenue and facilitate human decision-making. 
In this study, we build a sophisticated multimodal neural network that predicts box offices by grounding crowdsourced descriptive keywords of each movie in the visual information of the movie posters, thereby enhancing the learned keyword representations, resulting in a substantial reduction of 14.5\% in box-office prediction error.
The advanced revenue prediction model enables the analysis of the commercial viability of ``copycat movies," or movies with substantial similarity to successful movies released recently. We do so by computing the influence of copycat features in box-office prediction. We find a positive relationship between copycat status and movie revenue. However, this effect diminishes when the number of similar movies and the similarity of their content increase. Overall, our work develops sophisticated deep learning tools for studying the movie industry and provides valuable business insight.
\end{abstract}


\section{Introduction}
\renewcommand{\thefootnote}{\fnsymbol{footnote}}
\footnotetext[1]{This article is a significantly revised and expanded version of a paper originally presented at the IEEE ICME 2023, held on July 10th in Brisbane.}
\footnotetext[2]{© 2025 IEEE. Personal use of this material is permitted. Permission from IEEE must be obtained for all other uses, in any current or future media, including reprinting/republishing this material for advertising or promotional purposes, creating new collective works, for resale or redistribution to servers or lists, or reuse of any copyrighted component of this work in other works.
}
\renewcommand{\thefootnote}{\arabic{footnote}}

\IEEEPARstart{M}{ovie} is a preeminent form of art in the modern era, but the business side of movie production is often less than glamorous. Statistics \cite{pan2010} show that box-office revenues have a long-tailed and bimodal distribution, with a small number of movies taking the lion's share of profits.  
The skewed distribution of movie revenues and ever-increasing production costs mean that movie production carries significant risks. 
The exorbitant risks underscore the importance of understanding market dynamics and accurately predicting box-office revenues \cite{mckenzie2023economics,behrens2021leveraging}. 

One risk mitigation strategy that has become increasingly popular in recent years is to produce movies similar to recent successes \cite{eliashberg2006motion}. This strategy includes sequels, franchise movies set in the same fictional world, such as the Marvel Cinematic Universe movies and Justice League movies, and movies featuring similar story themes. 

However, the failure of superhero sequels like \emph{The Flash} (2023), which resulted in losses exceeding \$200 million, may have cast doubt on the effectiveness of this strategy. Interestingly, the literature on the sequel strategy contains inconsistent findings. Some research finds that sequel movies benefit from high name recognition that attracts a large audience \cite{ravid2004managerial, basuroy2006empirical, belvaux2021prevision}. Others suggest that repetition of similar content can lead to audience saturation that negatively affects the box office \cite{dhar2012long, sood2006brand}. In addition, most existing studies are limited by small datasets and often analyze only the first sequel movie. Thus, there remains much room for further examination.

In this paper, we study the task of predicting movie box-office performance based on an extended definition of such similar movies --- \emph{copycat} movies that share a common story theme. 
We argue that the franchise/sequel strategy should be examined from a broader perspective, specifically through the lens of content similarity. For instance, following the success of \emph{The Hunger Games} (2012), movies with similar themes, such as \emph{The Maze Runner} (2014) and \emph{Divergent} (2014), were quickly pushed to the market. Although these movies are not exact sequels, nor do they directly share the same fictional universe, they do share common characteristics, such as strange dystopian worlds, young adults working as a team, and strong female lead characters. Hence, we coin the term \emph{copycat} to describe this phenomenon. Note that a sequel or franchise can be categorized as a copycat movie only if it covers similar themes in terms of story content.



\begin{table}[!t]
    \centering  
    \small
    \caption{Examples of  user-generated keywords from TMDB.}\label{tab:keywords_result}
    \begin{tabular}{@{}c@{}}  
        \toprule
        \makecell{action, criminology, fbi, psycho, aircraft, robot\\ love, hate, high school, father-daughter relationship, \\ paris france,  kingdom, based on novel or book}\\
        \bottomrule
    \end{tabular}
\end{table}

We propose sophisticated machine learning techniques to overcome several methodological challenges that arise when studying copycat movies defined based on story/content themes. The first challenge is to quantify content similarity and identify copycats. We utilize user-generated movie keywords from The Movie Database (TMDB)\footnote{\url{www.themoviedb.org}} as proxies for movie content themes. Table \ref{tab:keywords_result} shows some example keywords.
Compared to traditional genre categories, these keywords provide a finer content categorization, including topic, plot, emotion of the plot, and even source-related information.\footnote{More details are in the keyword contribution guide located at \url{https://www.themoviedb.org/bible/movie}} The keyword data, which is used to define copycat movies, is highly idiosyncratic, featuring many near-synonyms and missing occurrences. To deal with these issues, we cluster the keywords using word embeddings learned from textual data and movie-keyword co-occurrence statistics, producing informative keyword clusters that allow the identification of similarly themed movies.  

The second challenge is building a strong model that predicts box-office revenue accurately while accounting for the impact of the copycat strategy. To this end, we propose a multimodal pretraining strategy that grounds content keywords in the visual imagery of movie posters. 
Our approach is motivated by the observation that the meaning of keywords in the movie context may be subtly different from the meaning in daily usage, which is captured by pretraining on regular text corpora. For example, the movie genre \emph{action} may be associated with explosions, car chases, or martial arts, substantially deviating from its dictionary definition. The keyword \emph{robot} typically refers to humanoid robots in science fiction or animated movies rather than robotic arms on an assembly line. The proposed pretraining strategy incorporates movie posters into the representation learning of keywords, producing pretrained network parameters that are conducive to movie revenue prediction.



Empirical results reveal the effectiveness of the proposed technical improvements. Visual grounding pretraining reduces test error by 14.5\% compared to a strong random forest baseline and by 4\% compared to a pretrained \text{BERT} model with the same number of parameters. Our results are comparable to those obtained from the advanced multimodal large language model backbone, LLaVA-7B \cite{liu2024visual_llava}, and slightly outperform those achieved with the CLIP \cite{zhang2024longclip} backbone. On top of this strong model, we show that incorporating movie content features indeed improves the box-office revenue prediction. Specifically, we show movie content variables (e.g., keywords, copycat-related variables) significantly reduce the prediction error, achieving up to 6.9\% improvement over the baseline model. 

With this paper, we make the following contributions. The first three are technical contributions, whereas the last one provides business insight.  
\begin{enumerate}[label=\textbullet]
    \item We construct a multimodal dataset \footnote{https://github.com/jdsannchao/MOVIE-BOX-OFFICE-PREDICTION} of 35,794 movies, consisting of textual information and posters, to facilitate box-office revenue prediction.
    \item We propose a method to identify \emph{copycat} movies that contain non-original content, using user-generated keywords from our dataset as descriptors of the movie content. To model keywords effectively, we use co-occurrence-based embeddings to summarize the long-tailed keyword list into a concise set.
    \item We propose a two-stage training procedure for the box-office prediction task, including a self-supervised stage that learns informative keyword representations using visual grounding and masked field prediction, and a second stage of finetuning on box-office data.
    \item In order to provide insight for business decision-making, we identify the key features influencing box-office revenue and quantify the effects of the \emph{copycat} strategy. Our study indicates that the success of copycat movies depends on their relative timing with other copycat movies.
\end{enumerate}

\section{Related Work}

\subsection{Predicting Movie Success} 
Previous works have attempted to predict various indicators of a movie's commercial and artistic success, including the box office \cite{ kim2023does,apala2013prediction_lash1,hur2016box}, return on investment \cite{eliashberg2014script1_lash11,lash2016earlyprediction_lash}, ratings \cite{cizmeci2018predicting_sheet8}, and awards (or nominations) \cite{boccardelli2008critical_lash5}. More recently, with the advancement of Machine Learning, applications of deep networks in such tasks have begun to gain attention \cite{quader2017performance_sheet10, antipov2017box_sheet12, zhou2018, kim2019prediction, shafaei2019starpower1_sheet1}.

Aside from commonly adopted numeral features, available textual features include movie reviews and movie content. Previous research has mainly focused on creating topic distance matrices and conducting sentiment analysis. 
Studies constructing topic distance matrices have utilized the Bag-of-Words approach \cite{eliashberg2014script1_lash11, lash2016earlyprediction_lash}, Latent Dirichlet Allocation \cite{kim2023does}, and the pretrained fastText word embeddings \cite{shafaei2019starpower1_sheet1}. Sentiment analysis has also been used to analyze audience reviews, as word of mouth is an important factor for box-office prediction \cite{apala2013prediction_lash1, hur2016box, yu2010mining}. Using movie content information is crucial for other downstream tasks, such as movie question answering and recommendations. The methods used include encoding subtitles with skip-gram \cite{yuan2020adversarial} and LSTM \cite{zhao2017social} or transforming subtitles into semantic descriptions \cite{liang2012script, kurzhals2016visual}.

Unlike the aforementioned approaches, our approach to analyze movie content builds upon the pretrained embeddings by incorporating visual poster information through self-supervised pretraining.
To our knowledge, the only prior work integrating poster information with text using deep learning models for box-office prediction is \cite{zhou2018}, which employs an evolutionary algorithm to select the optimal convolutional neural network (CNN) architecture, focusing on the layer at which visual features should be fused. In contrast, our approach does not directly integrate visual features into the box-office prediction task. Instead, we improve the pretraining of text embeddings through a visual grounding method.

\subsection{The Copycat Effect} \label{sec:copycat_intro}

A copycat commonly refers to a product that heavily imitates the functionality, design, and content of existing products \cite{naik1998planning, van2012consumer, wang2018copycats} and can be found in various industries, such as consumer goods \cite{naik1998planning,van2012consumer} and mobile apps \cite{wang2018copycats}. In the movie industry, copycats can be considered as those works that closely resemble previous blockbusters in terms of content, just like the previous example of \emph{The Divergent} (2014), which can be seen as a copycat of \emph{The Hunger Games} (2012).

Studies on the effect of copycat products over original products on profit yield have sparked debates in various industries. Copycat mobile games, for instance, have generated higher revenues than originals due to similar pricing and more downloads \cite{wang2018copycats}. However, in advertising, style imitation makes it harder for products to stand out as competition for consumer attention intensifies \cite{naik1998planning}.  In the movie industry, existing research focuses on the profitability of sequel movies \cite{sood2006brand, dhar2012long, ravid2004managerial, basuroy2006empirical, belvaux2021prevision}. However, our definition of ``copycats" is broader than ``sequels'' and not limited to the same cast team or the same story background.

We define the concept of ``copycat movies" and conduct a fine-grained analysis on a large-scale dataset. We find that copycat movies can generate higher box-office revenue in the early stages due to their similarity to blockbusters. However, over time, the worn-out effect may become more prominent and lead to audience fatigue and boredom \cite{naik1998planning}, which in turn results in a decline in box-office revenue.

\subsection{Self-supervised Multimodal Pretraining} 

The success of pretrained textual models such as \text{BERT} \cite{devlin2018BERT} has inspired a series of pretrained multimodal models \cite{lu2019vilBERT, su2019vl, tan2019lxmert, huang2020pixelBERT}, often leveraging the masked language modelling (MLM) objective. Similar to a denoising autoencoder, the MLM objective trains the model to predict masked portions of the input. This seemingly simple training technique has demonstrated effectiveness across various downstream applications. 
Another line of work, such as CLIP \cite{radford2021learning_clip} and BLIP \cite{li2022blip}, trains the network to distinguish between correct and incorrect image-text pairs. 

The symbol grounding problem \cite{Harnad_1990}, a classic problem in cognitive science, concerns how words can gain their meaning as pointers to other concepts and objects. Computationally grounding textual tokens in visual images has demonstrated success in some applications \cite{kiros-etal-2018-illustrative, tan-bansal-2020-vokenization, ZhangLiSai2021, YujieLu-Imagination-2022,yang2022z, liu-etal-2022-things}. In this work, we use movie posters as a visual grounding source for textual tokens -- keywords. A movie poster is a widely used visual medium to promote a movie long before its release. Thus, we ground the tokens using objects from a single poster, allowing multiple associations between tokens and objects. 
Unlike previous works that retrieve or generate relevant images for the textual descriptions, in our task, the correspondences between the keywords and the poster are not known \emph{a priori} and must be discovered in a multi-instance manner.

\section{Box-office Data Collection and Prediction Network} 
\label{sec:method}

Here we define the problem, outline data collection and feature engineering, and present our network architecture with self-supervised pretraining.

\subsection{Problem Definition}

Accurately predicting a movie's box-office revenue $y_i$ is a complex task influenced by multiple factors, including content, release timing, cast, and marketing. To tackle this challenge, we propose deep learning models with self-supervised training that leverage both textual and visual features to predict revenue. Formally, our goal is to learn a predictive function $f$ that maps movie information $m_i$ to its corresponding box-office revenue $y_i$.

\subsection{Data Collection}   
To set a public benchmark for box-office prediction, we collect metadata for 35,794 movies from TMDB, spanning from 1920 to 2022. The total box-office revenue for each movie during its release period is retrieved from IMDbPro.com.




\subsection{Keyword Content Features}\label{sec:keywords}
 
The dataset includes user-generated keywords for each movie, yielding 7,700 unique keywords for 35,794 movies. We observe many rare words and near-synonyms that may hinder learning. Rare keywords lack training data, which may prevent accurate embedding learning. Synonyms and near-synonyms would force the model to learn dissimilar embeddings for words with similar meanings. We also noticed that some less popular movies have limited views, resulting in missing keywords. These inadequate keywords fail to fully reflect the content of the movie. To tackle these issues, we utilize keyword lexical similarity and co-occurrence statistics to create keyword clusters, effectively tackling both the issues of near-synonyms and missing keywords. 

We use 300-dimensional embeddings computed by fastText \cite{bojanowski2016enriching} to represent the lexical information. To capture co-occurrence statistics, we employ the term-frequency inverse-document-frequency (TF-IDF) matrix and the {\sc Leporid} \cite{zhang2021initialization} technique, which is a regularized Laplacian Eigenmap \cite{Luxburg2007ATO}. 
We extract the first 50 dimensions of eigenvectors as the embedding to represent a keyword. The final representation is the 350-dimensional concatenation of the two vectors. We then perform average-link agglomerative clustering and use the resultant 1,140 keyword clusters as features of movies. We show some examples of the clustering results in Table \ref{tab:clusters_result}. We discuss the effect of keyword clusters instead of using raw keywords later in \S{\ref{sec:experimental_results} and Table \ref{tab:models_compare_clustering}.}

\begin{table}[t]
    \centering  
    \small
    \caption{Examples of the clustering results.}\label{tab:clusters_result}
    \vspace*{2mm}
    \begin{tabular}{|c|c|}  
        \hline
        Cluster Label & Elements \\
        \hline
        love-hate & \makecell[l]{`love', `loved', `hate', `unhappy',\\`waiting', `happy', `grateful', `lucky',\\`expecting', `loving'}\\
        \hline
        superhero & \makecell[l]{`superhero', `villainess', `villain', `symbiote',\\ `sidekick', `superhuman', `teamup', `nemesis',\\`superheroes', `supervillain'}\\
        \hline
        psychopathy & \makecell[l]{`psycho', `psychotic', `pyromaniac',\\ `psychopathic', `homicidal', `deranged'}\\
        \hline        
    \end{tabular}
\end{table}

\subsection{Copycat Features} 

Copycat movies are driven by the desire to replicate the success of a blockbuster movie. Therefore, to identify copycat movies, we first define blockbusters as those generating at least \$10 million in revenue and having a revenue-to-budget (return on investment) ratio of at least 3. This resulted in 2,486 blockbusters in our dataset. Next, we quantify the content similarity between each movie and a blockbuster as Jaccard similarity $J(A, B) = \frac{|A \cap B|}{|A \cup B|}$ over the keyword sets $A$ and $B$, where $A$ and $B$ denote the keyword clusters associated with the two movies, respectively. 

For each blockbuster movie, we keep the top 10 most similar movies released within ten years after as its \emph{copycats}. We keep the Jaccard similarity as a feature called \emph{copycat similarity}. If a movie is not considered as a copycat, its value is set to zero. This feature enables us to examine the relationship between a movie's revenue and its similarity to blockbusters. Additionally, we use the chronological order of the 10 copycat movies for each blockbuster as a feature called \emph{copycat rank}. If a movie is not a copycat, its copycat rank is zero. This feature allows us to examine the relationship between release time and the success of copycat movies. 



A related concept is franchise movies. These movies are typically set in the same fictional universe, such as the Marvel Cinematic Universe or the Alien Universe, and share the same fictional universe, the common artistic vision, a compatible and co-referencing storyline, and continuous marketing strategies. While these movies may not be highly similar in content, they still benefit from the advertising and reputation of previous movies. Controlling for this feature allows us to more accurately observe the copycat effect. The label indicating whether a movie belongs to a franchise or not is sourced from the TMDB dataset, which we adopt and refer to as \emph{franchise}. Of the 35,794 movies, we identify 4,211 copycats, 956 of which are also franchise movies. The statistics on copycats and franchises can be found in Figure \ref{fig:distribution_copycat} and Table \ref{tab:copycat_stats} in the Supplementary Material.

\subsection{Other Common Movie Features} \label{section3.1}
We also conducted feature engineering across four main movie feature categories: (1) Basic information, such as genres and MPAA ratings; (2) Production and marketing information, including the production and distribution companies as well as the budget; (3) Release timing and market competition, where competition refers to the number of other movies in the same genre released around the same time; and (4) The box office influence of the cast and crew, termed ``Star Power," where we analyze the number of previously movies released by cast and crew to access their experience, and calculate the average box office revenue of their past movies to measure their profitability. The details can be found in \S\ref{supp:features_eng} and Table \ref{tab:var_desc} of the Supplementary Material.

\subsection{Self-supervised Pretraining}
\begin{figure*}[ht]
\centering
\includegraphics[scale = 0.5]{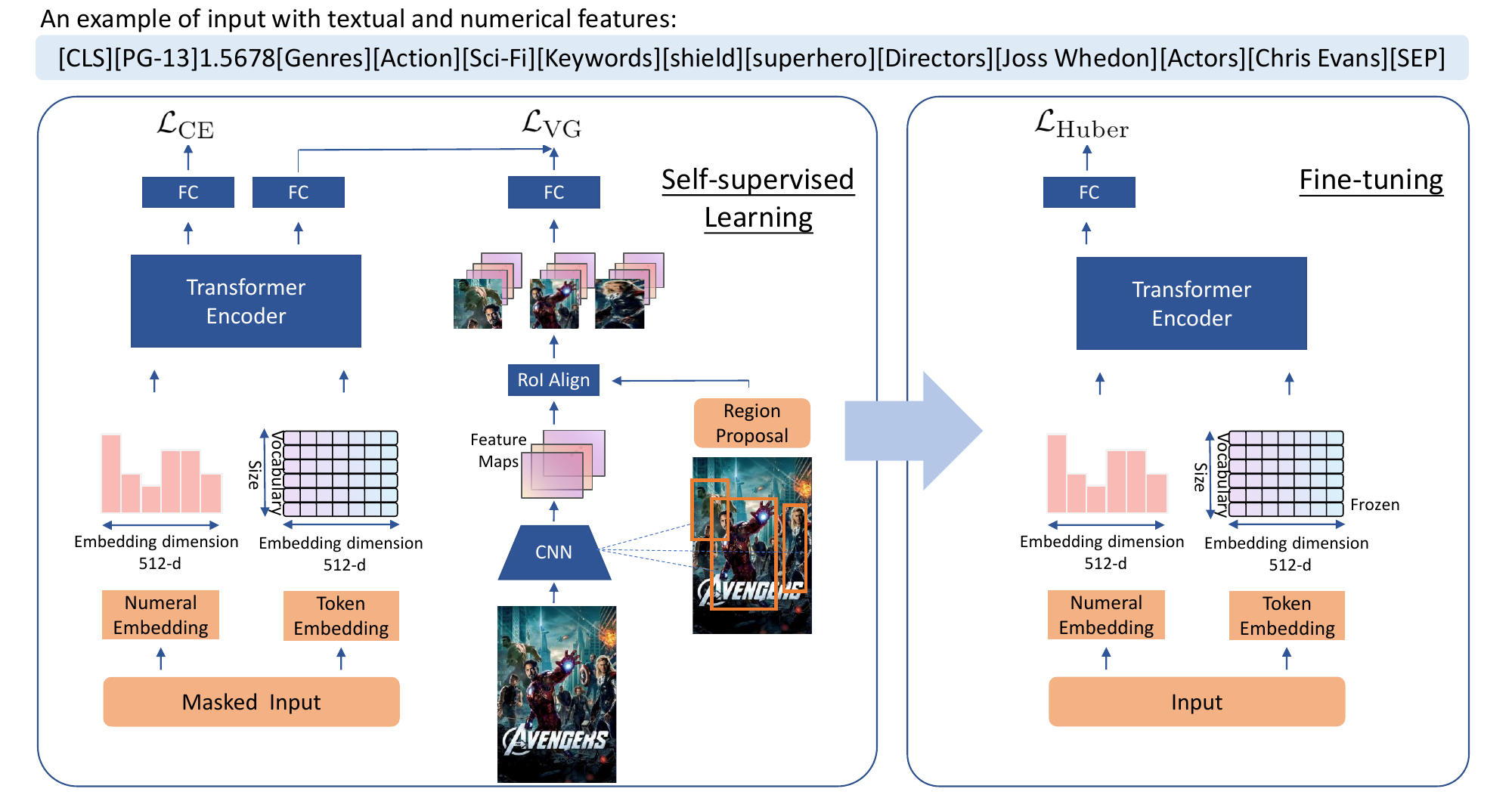}
\caption{The overall pipeline of self-supervised pretraining and finetuning on the box-office prediction task. The token embeddings are frozen during finetuning.}
\label{fig:figures_for_paper_overview}
\end{figure*}

Figure \ref{fig:figures_for_paper_overview} shows the overall pipeline. In the first stage, we pretrain a Transformer network on the MLM and visual grounding objectives. Next, we freeze the token embeddings and finetune the network on box-office prediction. 

\noindent \textbf{Token Embedding and Numerical Embedding.} For each value in discrete features (i.e., a discrete token), we create an embedding vector whose values are learned from data. For real-valued features, we adopt prototype-based numeral embeddings \cite{jin2021numgpt}. Formally, the embedding function is formulated as $\operatorname{NE}(x): \mathbb{R} \rightarrow  \mathbb{R}^{D}$ that maps a real number $x$ to a $D$-dimensional vector with the component $\mathrm{NE}_i(x)=\exp \left(-\frac{\left\|x-q_i\right\|_2}{\sigma^2}\right)$,  where $\{q_i\}^{D-1}_{i=0}$ are $D$ evenly spaced numbers over a specified interval, e.g., $[-10,10]$. Before applying the numerical embedding function, we normalize the values using logarithm or min-max normalization, depending on whether or not the feature has a long-tail distribution. 

\noindent \textbf{Masked Field Prediction.} 
We adopt a pretraining objective similar to the masked language modeling task, which has been shown to be effective for natural language understanding \cite{devlin2018BERT} and multimodal understanding \cite{lu2019vilBERT}. 
We randomly mask one token from each group of input features—genres, keywords, director/writer names, and actor names—and train the network to predict the missing token. The prediction is formulated as cross-entropy losses, which we denote as $\mathcal{L}_\textrm{CE}$. By training the network to predict missing fields, we encourage the network to learn correlations between inputs, which could mitigate data scarcity issues. 

\noindent \textbf{Structured Visual Grounding.}
Understanding keywords, which reflect movie content, is a challenging task. We propose to ground the keywords in the visual information from movie posters by contrastive learning that encourages high similarity between a poster and the corresponding content keywords and suppresses the similarity between incorrectly paired posters and keywords. First, we perform object detection on the poster with an off-the-shelf network. We denote the extracted object features from the $i^{\text{th}}$ movie as $\mathcal{Z}_i = \{\boldsymbol{z}_{m}\}^{M}_{m=1}$, $M$ refers to the number of objects. Note that we use the subscript $i$ to denote the movie index. We also use contextualized embeddings of the keywords from the output of the Transformer network, denoted as $\mathcal{X}_i ={\{\boldsymbol{x}_{k}\}}^{K}_{k=1}$, $K$ refers to the number of keywords.  

\begin{figure}[!t]
    \centering
    \includegraphics[scale = 0.4]{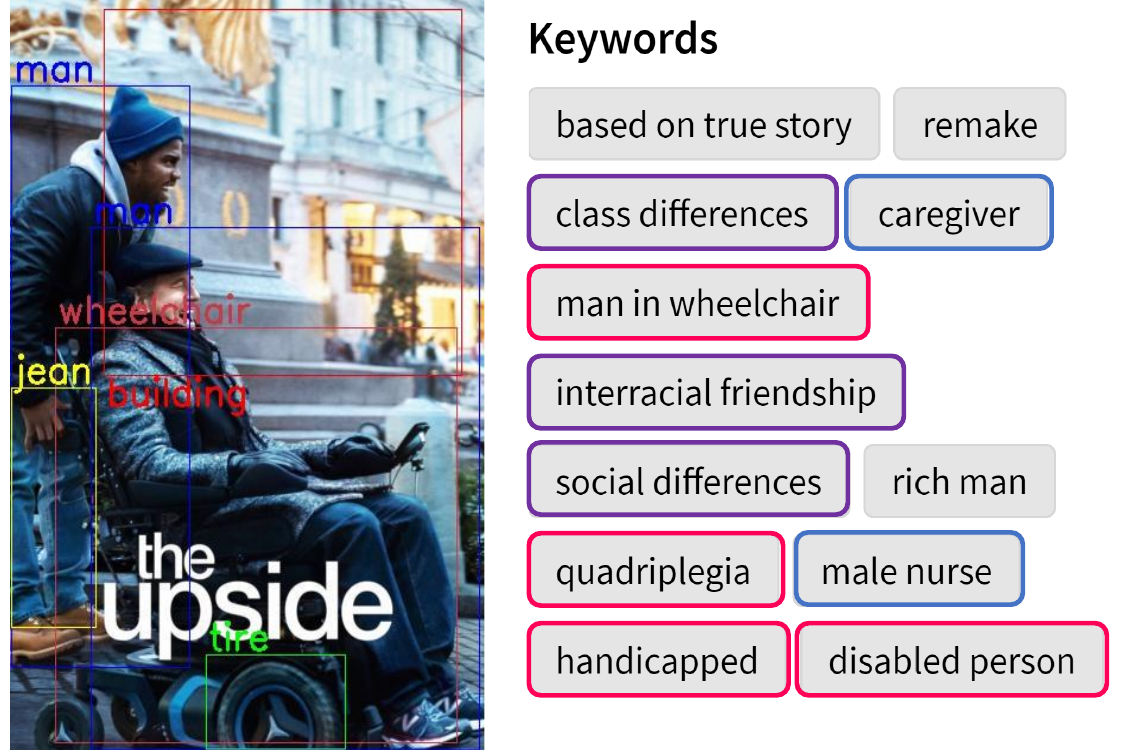}
    \caption{Multiple objects and keywords alignments for the movie \emph{The Upside} (2019)}
    \label{fig:theupside_keywords_links}
\end{figure}

We define the similarity between the poster and the keywords as 
\begin{equation}
\text{sim}(\mathcal{X}_i, \mathcal{Z}_i) = \sum_{(\boldsymbol{x}, \boldsymbol{z})\in \mathcal{X}_i\times \mathcal{Z}_i} \exp(\frac{\boldsymbol{x}^\top \boldsymbol{z}}{\|\boldsymbol{x}\|_2 \|\boldsymbol{z}\|_2}),
\end{equation}
where $\times$ denotes the Cartesian product and $\| \cdot \|_2$ denotes the L2 norm. Due to many-to-many relations between objects on the poster and the keywords, we follow \cite{miech2020end} to define the similarity as the sum of similarities of all possible pairs. To illustrate this, we show one example poster and the associated keywords in Figure \ref{fig:theupside_keywords_links}. Keywords in boxes of the same color belong to the same keyword cluster (e.g., ``quadriplegia'' and ``handicapped'' belong to the red cluster). One keyword cluster can match multiple objects, and one object may be grounded in multiple clusters. For instance, the cluster ``quadriplegia'' is grounded by the wheelchair, the tire, and the sitting man; the sitting man is related to the red and the purple clusters. 

With randomly sampled negative pairs $(i^\prime, j^\prime)$, we define the visual grounding (i.e., VG) loss, $\mathcal{L}_\text{VG}$, as

\begin{equation}
\mathcal{L}_\text{VG} = -\frac{1}{N} \sum_{i=1}^N \log \left(\frac{ \text{sim}(\mathcal{X}_i, \mathcal{Z}_i)}{\text{sim}(\mathcal{X}_i, \mathcal{Z}_i) + \sum_{(i^\prime, j^\prime)} \text{sim}(\mathcal{X}_{i^\prime}, \mathcal{Z}_{j^\prime})}\right)    
\end{equation}
where $N$ is the total number of movies in the training set.

\noindent

\subsection{Finetuning on Box-office Prediction} 
In the finetuning stage, we train the network to predict box-office revenues. We generate the prediction by feeding the average output from all input positions into a fully connected layer. Since revenues follow a long-tailed distribution, we take the base-10 logarithm of the revenue as the target value. To further mitigate the impact of outliers, we train the network using the Huber loss,
\begin{equation}
    \mathcal{L}_\textrm{Huber}= \begin{cases}0.5\left(y-\hat{y}\right)^2,     \text{ if}\left|y-\hat{y}\right|< 1  \\ 
    \left|y-\hat{y}\right|-0.5  , \text{ otherwise}\end{cases},
\end{equation}
where $y$ is the ground truth and $\hat{y}$ is the prediction. Huber loss uses an absolute value when $|y-\hat{y}| > 1$ and the square of the error when $|y-\hat{y}| \leq 1$. This makes it less sensitive to outliers compared to Mean Squared Error (MSE) and it is differentiable at 0 unlike Mean Absolute Error (MAE). It is widely used in computer vision (CV) object detection as Smooth L1 loss, such as in Fast R-CNN \cite{girshick2015fast}.

\section{Experimental Results}\label{sec:experimental_results}

\subsection{Setup}
We use stratified sampling to divide the data into train, validation, and test sets in the ratios of 70/10/20, using ``franchise movie'' as the label for stratification. Using the method in \S~\ref{sec:keywords}, we cluster 7,700 keywords into 1,414 clusters. The number of clusters is tuned on the validation set.

\subsection{Baselines} 
We introduce three types of baseline models. First, we have the traditional machine learning model, Random Forest (RF). We feed all numerical features and a subset of categorical variables—those with a small number of possible classes—into RF, as one-hot encoding of all categorical features would result in excessively high dimensionality. Second, we introduce pretrained \text{BERT} models of small. This baseline does not involve MLM or VG pretraining and we only finetune it on box-office prediction. To mimic the classic \text{BERT} input, we concatenate all input tokens into one sentence while rounding numeral features to one decimal point and then apply the \text{BERT} tokenizer. Third, we select advanced multimodal models, CLIP and LLaVA-1.5-7B \cite{liu2024visual_llava}. We choose Long-CLIP \cite{zhang2024longclip} because it extends the vanilla CLIP input length to 248 tokens, suitable for accommodating our input data. For the CLIP baseline, we use only textual information as input. For the LLaVA-1.5 baseline, we conduct separate tests with and without the image poster as input. We freeze both models' parameters and extract the \texttt{[CLS]} token output from the final layer. On top of this, we train linear projector layers for box-office prediction for each model. The size of this layer remains consistent across all experiments, except for the RF as it is not applicable.

\subsection{Implementation Details}
Our model consists of a 4-layer Transformer with a dimension $d_{\text{model}}=512$, fully connected layer dimension $d_{fc}=512$, and attention heads $H=4$. The architecture is the same as $\text{BERT}_\text{small}$. During visual grounding pretraining, we randomly select up to 6 keywords per movie and up to 20 objects per poster to compute the similarity. The feature map of each object has $2048$ channels and $4 \times4$ pixels, which is the output from VinVL \cite{zhang2021vinvl}, the off-the-shelf object detection model, after ROI Align \cite{he2017maskrcnn} and an adaptive average pooling layer. The feature map is flattened spatially and then linearly projected to $\mathrm{R}^ {d_{model}}$, where $d_{model} = 512$. 

We use a batch size of 2,048 when pretraining the model under the MLM objective and reduce the size to 326 when adding the VG objective. The learning rate is 3e-4. We used Adam optimizer with a weight decay equal to 1e-4. During the finetuning stage, we freeze the model, use the last layer output of \texttt{[CLS]} token, and train a linear projector (similar to other Transformer baselines). We search for the best performance on the validation set in the combinations of learning rate in [1e-3, 3e-4, 1e-4] and batch size in [328, 512, 1024].

\subsection{Model Performance}
In Table \ref{tab:models_compare}, we report Huber loss on the test set from the box-office prediction task for all models, as well as their performance relative to $\text{BERT}_\text{small}$. Pretrained \text{BERT} models easily outperform the RF but are inferior to the MLM and VG pretraining. Our best model outperformed $\text{BERT}_\text{small}$ by 14.5\%. Notably,  including VG pretraining resulted in a significant improvement on top of masked MLM. The addition of VG reduced the loss from 0.3102 to 0.3037. 

Compared to other advanced multimodal baselines, first, the result shows that Long-CLIP (Huber Loss = 0.3213) underperforms compared to our proposed pre-training method (Huber loss = 0.3037). Although CLIP is pre-trained on a general multimodal corpus, our method is self-pretrained on movie context data with visual grounding enhancement. By aligning the movie content representation (keywords) with movie context (poster), our approach achieves better performance with a three times smaller parameter size. 

Then, we tested LLaVA without poster input, following a similar evaluation process as with CLIP. The resulting performance was Huber Loss = 0.3092, slightly worse than our model (Huber Loss = 0.3037). We also tested LLaVA with poster input, which improved the performance to 0.3028, slightly surpassing our model. However, given the vast difference in model size, number of parameters (LLaVA: 7B vs. Ours: 161M), and pre-training data, the effectiveness of our model remains well-validated. 

Furthermore, since LLaVA’s training data ends in March 2023\footnote{\url
{https://lmsys.org/blog/2023-03-30-vicuna/}}, while our training data ends in 2022, we collected additional box-office data from April 2023 to December 2024 to serve as a new test set (sample size = 789). On this dataset, our model outperformed LLaVA (ours: 0.7272 vs. LLaVA: 0.7406), as shown in Table \ref{tab:new_test_compare}. 

To make the loss more interpretable based on the original scale of the box office, we calculated the Mean Absolute Percentage Error (MAPE) using the following formula: 

\begin{equation}
\text{MAPE} = \frac{1}{N} \sum_{i=1}^{N} \left| \frac{\hat{y}_{i} - y_{i}} {y_{i}} \right|
\end{equation}

We then divided the movies into four buckets based on their total box-office revenue. Table \ref{tab:MAPE} shows that our predictions are more accurate for high box-office movies in general. For a movie with an actual box-office revenue of 1 billion (1B), a MAPE of 0.52 indicates that the predicted revenue typically falls between 0.48 billion and 1.52 billion, demonstrating a good level of prediction accuracy.

\begin{table}[t]
\small
\centering
    \caption{Comparison of the test Huber loss between models. A lower loss indicates higher prediction accuracy. }\label{tab:models_compare}
\vspace{2mm}
\begin{tabular}{lc}
\toprule  
\makecell[c]{Model} &
  \makecell[c]{ Test Huber Loss $\downarrow$ \\\scriptsize{(\%  improve. over baseline)}}\\
\midrule   
Random Forest (RF)& $0.3677$\;${\color{Red}{(+3.5\%)}}$\,\;\\
\midrule
Standard $\text{BERT}_\text{small}$ & $0.3553$\;\text{(baseline)} ~\\
\midrule
\text{Long-CLIP} init.  & $0.3213$\;${\color{Green}{(-9.6\%)}}$\,\;\\
\text{LLaVA-1.5-7B} w/o Poster init.  & $0.3092$\;${\color{Green}{(-13.0\%)}}$\\
\text{LLaVA-1.5-7B with Poster} init.  & $0.3028$\;${\color{Green}{(-14.8\%)}}$ \\
\midrule
(Ours) \text{BERT} embeddings init.  & $0.3137$\;${\color{Green}{(-11.7\%)}}$\\
 + MLM pretraining   & $0.3102$\;${\color{Green}{(-12.7\%)}}$ \\
 + MLM\&VG pretraining & $0.3037$\;${\color{Green}{(-14.5\%)}}$ \\
\bottomrule
\end{tabular}
\end{table}

\begin{table}[t]
\small
\centering
    \caption{Comparison of the test Huber loss between LLaVA and our model with VG pretraining on movies released after 2023 Mar.  }\label{tab:new_test_compare}
\vspace{2mm}
\begin{tabular}{lc}
\toprule  
\multicolumn{1}{c}{Model} &
  \makecell[c]{ Test Huber Loss $\downarrow$ \\\scriptsize{(\%  improve. over baseline)}}\\
\midrule   
\text{LLaVA-1.5-7B with Poster} init.  & $0.7406$\;\text{(baseline)} ~\\
 \midrule
\text{BERT} embeddings init.  &\\
 + MLM\&VG pretraining & $0.7272$\;${\color{Green}{(-1.8\%)}}$ \, \\
\bottomrule
\end{tabular}
\end{table}

\begin{table}[t]
\small
\centering
    \caption{MAPE on box-office revenue after reversing the log10 transformation. Movies are categorized into four revenue buckets.}\label{tab:MAPE}
\vspace{2mm}
\begin{tabular}{lcc}
\toprule  
Revenue Bucket \;\;&  MAPE $\downarrow$ & Movie Count\\
\midrule  
\textless{}\$1M  & 46.582 & 3,698\\
\$1M-\$100M & 0.934 & 3,161  \\
\$100M-\$1B & 0.571 & 326  \\
\textgreater{}\$1B & 0.522 & 11  \\
\bottomrule
\end{tabular}
\end{table}

\subsection{Ablation Study}
This subsection discusses how different settings can improve the model performance. The keywords used in models have been clustered using the clustering method introduced in \S \ref{sec:keywords} otherwise specified.

\begin{table}[t]
\small
\centering
\caption{Comparison between the same model before and after incorporating copycat-related variables.}\label{tab:models_compare_copycat}
\vspace{2mm}
\begin{tabular}{lcc}
\toprule  
\multicolumn{1}{c}{Model} & \multicolumn{1}{c}{\makecell[c]{Copycat-related\\Variables}} &
  \makecell[c]{Test Huber Loss  $\downarrow$  \\\scriptsize{(\% performance difference)}}\\
 \toprule   
\multirow{2}{*}{\text{BERT} embeddings init.}                      & \ding{55} & $0.3207$\qquad\qquad\qquad\\
                                                            & $\checkmark$ & $0.3137$\;${\color{Green}{(-2.2\%)}}$ \\
\midrule
\multirow{2}{*}{+ MLM pretraining}    & \ding{55} & $0.3163$\qquad\qquad\qquad\\
                                                            & $\checkmark$ & $0.3102$\;${\color{Green}{(-1.9\%)}}$ \\
\midrule
\multirow{2}{*}{+ MLM\&VG pretraining} & \ding{55} & $0.3094$\qquad\qquad\qquad \\
                                                            & $\checkmark$ & $0.3037$\;${\color{Green}{(-1.8\%)}}$ \\
\bottomrule
\end{tabular}
\end{table}

\begin{table}[t]
\small
\centering
\caption{Comparison between the same model before and after incorporating content variables. }\label{tab:models_compare_content}
\vspace{2mm}
\begin{tabular}{lcc}
\toprule  
\multicolumn{1}{c}{Model} & \multicolumn{1}{c}{\makecell[c]{Copycat  Var.\\and Keywords}} &
  \makecell[c]{Test Huber Loss $\downarrow$\\ \scriptsize{(\% performance difference)}}\\
 \toprule   
\multirow{2}{*}{\text{BERT} embeddings init. }                  & \ding{55} & $0.3370$\qquad\qquad\qquad       \\
                                                         & $\checkmark$ & $0.3137\; {\color{Green}{(-6.9\%)}}$ \\
\midrule
\multirow{2}{*}{+ MLM pretraining} & \ding{55} & $0.3298$ \qquad\quad\;\:\,\\
                                                         &  $\checkmark$ & $0.3102\; {\color{Green}{(-5.9\%)}}$ \\
\bottomrule
\end{tabular}
\end{table}

\begin{table}[t]
\small
\centering
    \caption{The comparison between the same model before and after incorporating keywords clustering.}\label{tab:models_compare_clustering}
\vspace{2mm}
\begin{tabular}{lcc}

\toprule  
\multicolumn{1}{c}{Model} & \multicolumn{1}{c}{\makecell[c]{Keywords \\Clustering}} &
  \makecell[c]{Test Huber Loss $\downarrow$ \\\scriptsize{(\% performance difference)}}\\

 \toprule   

\multirow{2}{*}{Random embeddings init.}                      & \ding{55} & $0.3265$\;\;\;\;\;\;\;\;\;~~\;\;\;            \\
 & $\checkmark$ & $0.3290$\;${\color{Red}{(+0.8\%)}}$ \\

\midrule
\multirow{2}{*}{ + MLM pretraining}       & \ding{55} & $0.3133$ \;\;\;\;\;\;\;\;\;~~\;\:       \\
    & $\checkmark$ & $0.3109$\;${\color{Green}{(-0.8\%)}}$ \\

\midrule
\multirow{2}{*}{ + MLM\&VG pretraining} & \ding{55} & $0.3109$\;\;\;\;\;\;\;\;\;~~\;\;\;       \\
     & $\checkmark$ & $0.3070$\;${\color{Green}{(-1.3\%)}}$ \\

\toprule
\multirow{2}{*}{\text{BERT} embeddings init.}                      & \ding{55} & $0.3249$\;\;\;\;\;\;\;\;\;~~\;\;\;     \\
   & $\checkmark$ & $0.3137$\;${{\color{Green}{(-3.4\%)}}}$ \\

\midrule
\multirow{2}{*}{ + MLM pretraining}       & \ding{55} & $0.3226$\;\;\;\;\;\;\;\;\;~~\;\;\;         \\
      & $\checkmark$ & $0.3102$\;${\color{Green}{(-3.8\%)}}$ \\

\midrule
\multirow{2}{*}{ + MLM\&VG pretraining} & \ding{55} & $0.3182$ \;\;\;\;\;\;\;\;\;~~\;\:     \\
   & $\checkmark$ & $0.3037$\;${\color{Green}{(-4.6\%)}}$ \\
                                     
\bottomrule
\end{tabular}
\end{table}

\noindent \textbf{Effects of Content-related Variables.} We examine the impact of copycat variables and keyword variables on box-office revenue prediction.  Table \ref{tab:models_compare_copycat} shows that incorporating copycat-related variables leads to an average improvement of 2\%. Table \ref{tab:models_compare_content} shows that adding content keywords leads to an average performance improvement of 6.4\%. The table does not contain the VG loss, as it is impossible to apply the VG loss without content keywords.

\noindent \textbf{Effects of Keyword Clusters.} We verify the advantage of keyword clusters over raw keywords. Table \ref{tab:models_compare_clustering} compares models trained with keyword clusters (``Clustering") with those with raw keywords (``Keywords"). In most cases, keyword clusters provide performance gains, especially when pretrained \text{BERT} embeddings are used. One possible reason is that near-synonyms have similar \text{BERT} embeddings, making them difficult for model to differentiate, and clustering can alleviate this problem. 

Additionally, the keyword clustering provides greater benefit in VG pretraining models. The contrastive learning in VG pushes the embeddings of different keywords, $\bm{x}_i$ and $\bm{x}_j$, apart, which can be problematic if the two were in fact synonyms. In the keyword clustering setting, most synonyms are already clustered together, thereby mitigates this issue and enhances the discriminative power of contrastive learning. 


\section{Interpreting Feature Importance}\label{sec:explanatory}
Identifying variables that influence box office may facilitate the decision-making by movie producers and investors, such as deciding whether to produce original content or imitate blockbuster movies. In this section, we apply two interpretability methods, Attention Rollout \cite{rollout} and LIME~\cite{lime}, to estimate the impact of the features curated in \S \ref{sec:method}. 

\subsection{Attention Weights, Gradients and Attention Rollout}
We analyze the internal states of the model, specifically the attention weights, to assess the influence of a variable on the model's predictions. The attention weights reflect how the information of the tokens flows from input to output in the model -- tokens with higher attention weights contribute more to the representation of the tokens in the next layer. Thus, attention weights are used to provide plausible and meaningful interpretations \cite{wiegreffe2019attention,vashishth2019attention,vig2019visualizing}.

We adopt the technique Attention Rollout \cite{rollout,rollout_vit}. 
Consider a Transformer model with  $L$ layers. For the attention matrix $A_l$ at layer $l$, the weights in the $m$-th column indicate how much each token attends to the $m$-th token during token prediction. The weights in the $n$-th row show how much the $n$-th token attends to other tokens during the generation of the $n$-th token. The information flow from the $m$-th token at layer $l \shortminus 1$  to the $n$-th token at layer $l$  is calculated by the dot product between the $m$-th column of $A_{l \shortminus 1}$ and the $n$-th row of $A_l$. This operation sums the attention weights across all possible paths between the two tokens. Therefore, by performing matrix multiplication $A_lA_{l \shortminus 1}$, the resulting matrix represents the amount of information flowing to each token from every token in the previous layer. 
Notably, we choose to elementwise multiply $(*)$ the original attention matrix by its gradients $G_l$ with respect to the prediction loss, to highlight the attention weights that are important for minimizing the prediction loss. All negative values are then set to zero. The details are shown in Equation \ref{eq:attention_rollout}.

\begin{align}\label{eq:attention_rollout}
\begin{aligned}
A^{\prime}_l &= \max(G_l, 0) * A_l \\
\tilde{A}_l &=
    \begin{cases} 
      A^{\prime}_l & \text{if } l = 1 \\ 
      A^{\prime}_{l} \tilde{A}_{l \shortminus 1}  & \text{if } l > 1 
      \end{cases}
    \end{aligned}&
\end{align}

The $n$-th row of $\tilde{A}_L$ at the last layer represents the attention weights of each token, propagated from the input layer to the $n$-th token in the output layer.
We are only interested in the information flowed into the \texttt{[CLS]} token, as \texttt{[CLS]} token is the sole token used to predict the box-office. Thus, the first row $\vec{a} = \tilde{A}{_L}[0, :]$, which represents the influence of all tokens on the \texttt{[CLS]} token, is retained as the final result.


We next aggregate the vector $\vec{a}$ by averaging over the test set based on the occurrences of each variable value. 
For Genre and Month, we track the attention rollout values for each element value. 

\begin{figure}[t]
    \centering
    \includegraphics[scale = 0.49]{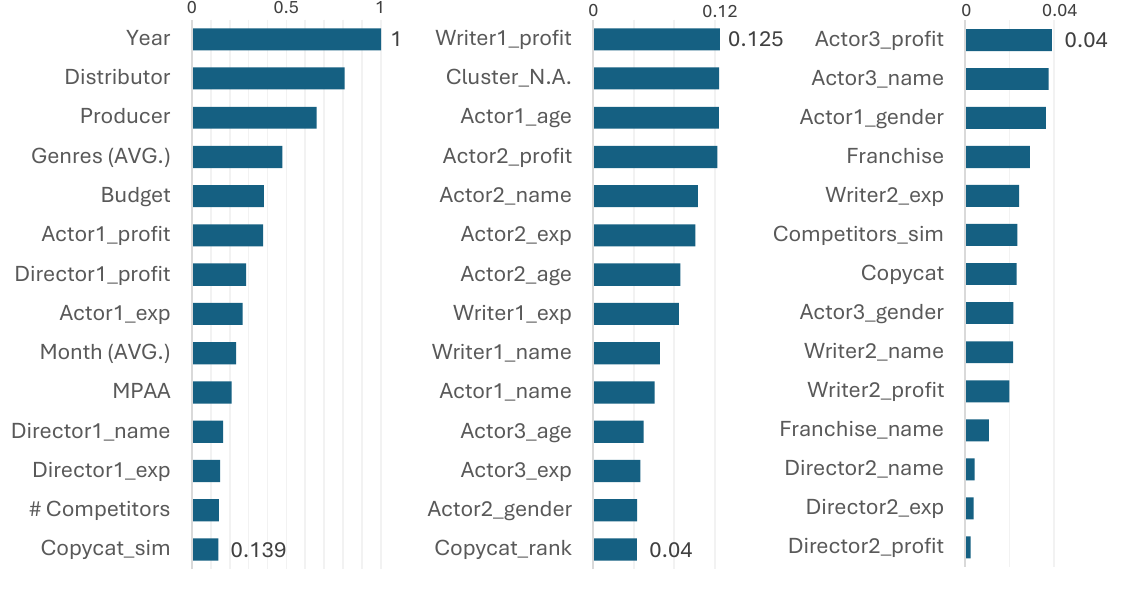}
    \caption{Attention rollout ranking. The values are normalized so that the maximum variable, \texttt{Year}, has a value of 1. The `1/2/3' designation for actors, directors, or writers indicates the order of prominence or billing. }
    \label{fig:attentionrollout_results}
\end{figure}


Figure \ref{fig:attentionrollout_results} shows the attention rollout ranking which is divided into three tiers from the highest to the lowest. The top-ranked features that have the greatest impact on box-office revenue are the movie's \texttt{Year}, \texttt{Distributor}, \texttt{Producer}, \texttt{Genres}  (on average) and \texttt{Budget}. 

Among the copycat-related features, the \texttt{copycat\_sim} is highly ranked. This suggests that the degree of a given movie's content being similar to other blockbuster movies is important in box-office prediction. 

While attention rollout shows the variable importance, it does not indicate whether each variable's impact is positive or negative. Furthermore, it does not quantify the magnitude of each variable's impact on the output (target). An additional interpretability technique is required to provide further insights.

\subsection{LIME}



The LIME (Local Interpretable Model-Agnostic Explanations) method \cite{lime} provides feature importance for a given data point. 
Specifically, we perform the following steps: Given the input data point, a $K$ dimensional vector, LIME first performs $N$ (here $N=5000$) random perturbations over all the dimensions, yielding $N$ synthetic data points, denoted as $X_s$. The perturbations follow the same distribution as the training data. Next, we gather model predictions on each synthetic data point, resulting in $N$ pseudo-labels, $\hat{y}_s \in \mathbb{R}^{N \times 1}$. Finally, LIME fits a Lasso regression model on $\langle X_s, \hat{y}_s \rangle$, and the $K$ regression coefficients represent the feature importance in the local model around the given input data point. After that, feature importance can be aggregated over many input data points.


Due to the sequential and heterogeneous nature of our data, which include both numerical variables and textual tokens, we perform the feature perturbation before embedding the input sequence. 
We perturb only a subset of features, including i) all numerical variables, and ii) most of the textual variables that have been simplified into categorical variables with no more than 12 categories. We do not perturb the extensive list of cast names and keywords. While these features are to the Transformer model, they are excluded from the Lasso regression, and hence will not produce LIME coefficients.

One thing to note is that since LIME provides a local explanation for each data point, for categorical variables, the LIME coefficient represents the influence of the specific variable value relative to other possible values. For instance, the ``Distributor" variable is categorized into 6 classes and we obtain 6 different LIME coefficients, one for each class.

\subsection{Results and Discussion}
\begin{figure*}[ht]
\includegraphics[scale = 0.47]{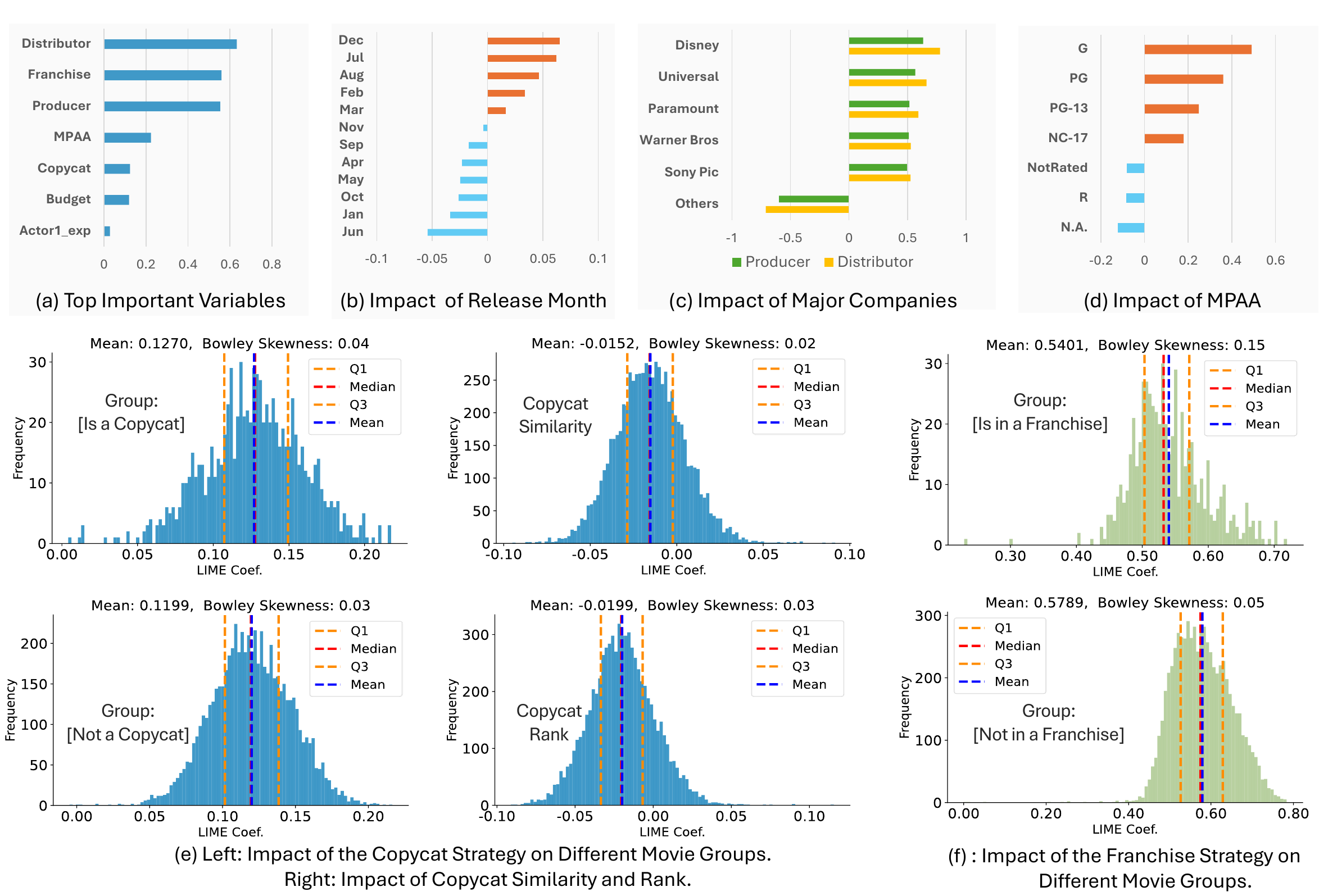}
\caption{The feature impacts using the LIME method. (a)-(d): Results for non-content related variables at the aggregation level. (e)-(f): The distribution plot of LIME coefficients for copycat-related variables (e) and the franchise variable (f).}
\label{fig:lime_results}
\end{figure*}
Figure \ref{fig:lime_results} displays the impact of variables as reflected by the LIME coefficients. 

\noindent\textbf{Importance of Commonly Used Features.} Figure \ref{fig:lime_results}(a) illustrates the top 7 most important features, as reflected by the largest absolute LIME coefficients. Except for \texttt{Budget} and \texttt{Actor1\_exp}, the rest are categorical features. Similar to attention rollout, LIME also identifies the \texttt{Producer} and \texttt{Distributor} features to have significant impacts.
As categorical features can take on different values, Figure \ref{fig:lime_results}(a) reflects the aggregated impact of each feature by averaging the LIME score for every possible values. We then show the impacts of those values in Figure \ref{fig:lime_results}(b)-(f). 

Next, we see a wide variation within \texttt{release\_month} in Figure \ref{fig:lime_results}(b). The coefficient $-0.058$ for June indicates that movies released in June are expected to have a lower box-office performance. In contrast, movies released in December are expected to see the increase in the box-office performance. 

Moreover, the marketing efforts by major movie companies have huge impacts on box office, as illustrated by Figure \ref{fig:lime_results}(c). These companies enjoy strong production and distribution capabilities. For example, being produced by Disney increases revenue by $0.6$ on average. The remaining four major companies are similar---a movie not produced or promoted by these companies will suffer revenue loss with coefficient of about $-0.6$.

Lastly, we have information about MPAA rating in Figure \ref{fig:lime_results}(d). As expected, a G rating shows the highest positive impact while an R rating and being unrated lower the revenue. Interestingly, NC-17 with sexual content turns out to have a positive impact.

\noindent\textbf{Importance of Copycat Features.} 
As shown in Figure \ref{fig:lime_results}(e) Top Left, the average LIME coefficient for the \emph{copycat} variable among the copycat movies is $0.1270$. Meanwhile, the LIME coefficient for changing a non-copycat movie to a copycat movie, other features being equal, is $0.1199$. Combining the above findings, being a copycat is associated with an approximately $0.12$ increase in $\log_{10}$ box-office revenue or roughly \$318,256 for a movie with revenue of 1 million dollars. 

Interestingly, as the LIME coefficient for \emph{copycat similarity} variable is negative ($\beta = -0.0152$, as shown in Figure \ref{fig:lime_results}(e) Top Right), we conclude that an increase of content similarity leads to a decrease in box-office revenue. Likewise, an increase in the \emph{copycat rank}, i.e., the more copycat movies stemming from the same blockbuster are released before the focal movie, has a negative impact on box-office revenue ($\beta$ = $-0.0199$, Figure 5(e) Bottom Right).

Taken together, we demonstrate an important insight: In general, movies categorized as copycat produce higher box-office revenue than original, non-copycat movies. However, too much similarity to the original blockbuster movie or the repetition of the same content over time across different copycat movies suppresses profitability.

\noindent\textbf{The Franchise Feature.}  Next, as shown in Figure \ref{fig:lime_results}(f) Top and Bottom, the LIME model consistently estimates an average positive impact of around $0.54$$\sim$$0.57$ on earnings for movies that are part of a franchise, whether they are originally franchise movies or non-franchise movies under LIME perturbation. Additionally, we find that LIME coefficients for being franchises have a large positive skew (under the Bowley skewness \cite{kenney1962mathematics} measurement), suggesting that franchise movies are likely to have extremely high box-office performances.

\noindent \textbf{Consistency between LIME and Attention.} By calculating the rank correlation (i.e., Spearman\textquotesingle s Coefficient) between the LIME and the attention rollout results, we get a value of $\rho = 0.6948$ (based on features in Figure \ref{fig:attentionrollout_results}, excluding textual information such as the names of casts). 
This indicates that the two sets of results are highly correlated with each other, suggesting our findings are not heavily reliant on the choice of the interpretability technique.

\section{Conclusion}

To conclude, our work has established a comprehensive process to predict movie box-office performance and provide valuable business insights for content management. The process begins with data collection, combining tokens, numbers, and visual information to represent the data. It then proceeds to the prediction task using the Transformer model through pretraining and self-supervision, and finally interprets the model's output. We believe our findings not only shed light on the importance of multimodal elements in box-office prediction but also demonstrate the interpretability of deep models in predictive tasks. This comprehensive approach offers significant implication for future research in this domain.

\section{Acknowledgments}
This research is supported, in part, by the RIE2025
Industry Alignment Fund – Industry Collaboration
Projects (IAF-ICP) (Award I2301E0026), administered
by A*STAR, as well as supported by Alibaba
Group and NTU Singapore through Alibaba-NTU
Global e-Sustainability CorpLab (ANGEL). The
research is also partially funded by the Nanyang Associate Professorship and National Research Foundation
Fellowship (NRFF13-2021-0006), Singapore.

\bibliographystyle{IEEEtran}
\bibliography{refs}

\vskip -2\baselineskip plus -1fil

\begin{IEEEbiography}
[{\includegraphics[width=0.8in,height=1.0in,clip,keepaspectratio]{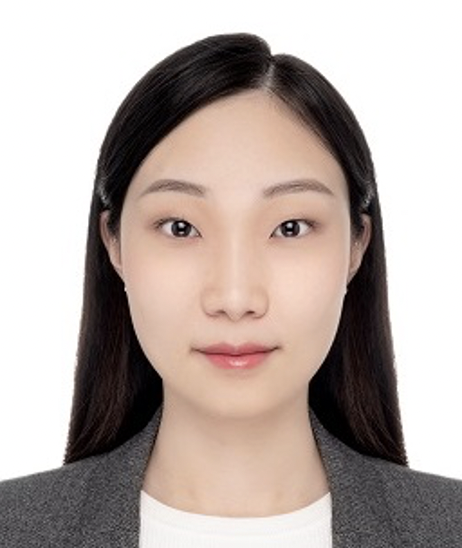}}]{Chao Qin}
is a Ph.D. student at the  College of Computing and Data Science (CCDS), Nanyang Technological University (NTU) in Singapore. She is exploring how stories come to life through words, images, and media. Her passion is uncovering the hidden meanings in these diverse storytelling forms.
\end{IEEEbiography}
\vskip -3\baselineskip plus -1fil
\begin{IEEEbiography}
[{\includegraphics[width=0.8in,height=1.0in,clip,keepaspectratio]{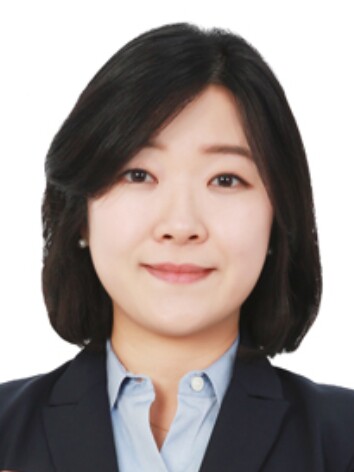}}]{Kim Eunsoo} is an Associate Professor of International Business at the University of Seoul. She holds a Ph.D. degree in Quantitative Marketing from the Stephen M. Ross School of Business at the University of Michigan, Ann Arbor. She was affiliated with NTU as an Assistant Professor of Marketing at Nanyang Business School. Her research focuses on quantitative modeling of user/firm-generated content and social influence, using econometric models, deep learning, and image/text analysis. 
\end{IEEEbiography}
\vskip -3\baselineskip plus -1fil
\begin{IEEEbiography}
[{\includegraphics[width=1in,height=1.25in,clip,keepaspectratio]{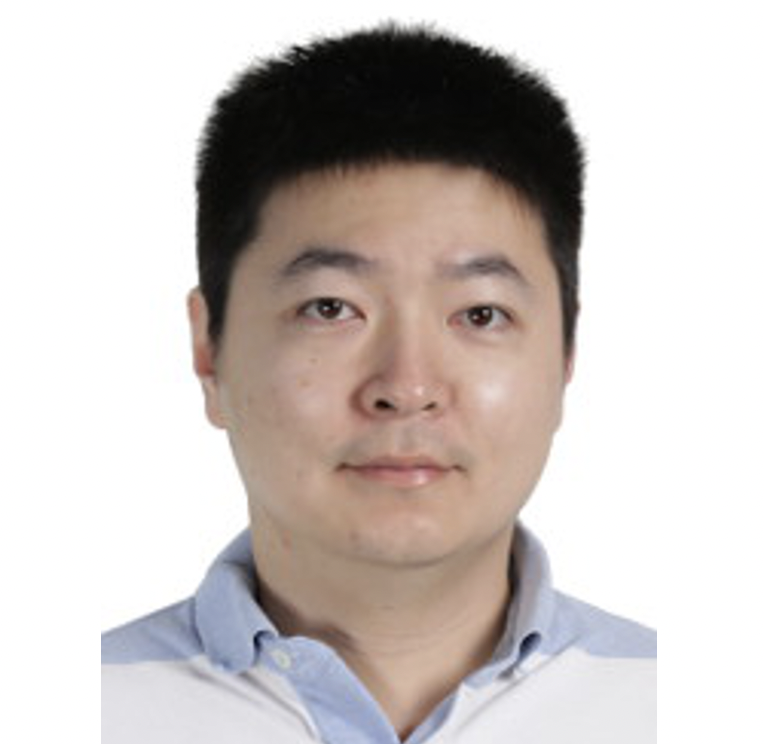}}]{Li Boyang} is a Nanyang Associate Professor at the College of Computing and Data Science, NTU. His research interests include computational narrative intelligence, multimodal learning, machine learning and data-centric AI. In 2021, he received the National Research Foundation Fellowship, a prestigious research award of 2.5M Singapore Dollars. He was a Senior Scientist at Baidu Research USA and a Research Scientist and Group Leader at Disney Research Pittsburgh. He received his Ph.D. degree from the Georgia Institute of Technology. 
\end{IEEEbiography}

\clearpage
\begin{appendices}
\setcounter{page}{1} 

\section{Authorship}
This is the supplementary material for the paper: ``Copycat vs. Original: Multi-modal Pretraining and Variable Importance in Box-office Prediction"
by Qin Chao, Eunsoo Kim, and Boyang Li. Inquiries regarding this paper can be directed to the authors at chao0009@ntu.edu.sg, eunsoo@uos.ac.kr, boyang.li@ntu.edu.sg. 

\section{Features  Engineering}\label{supp:features_eng}

\begin{figure}[h]
    \includegraphics[scale=0.44,left]{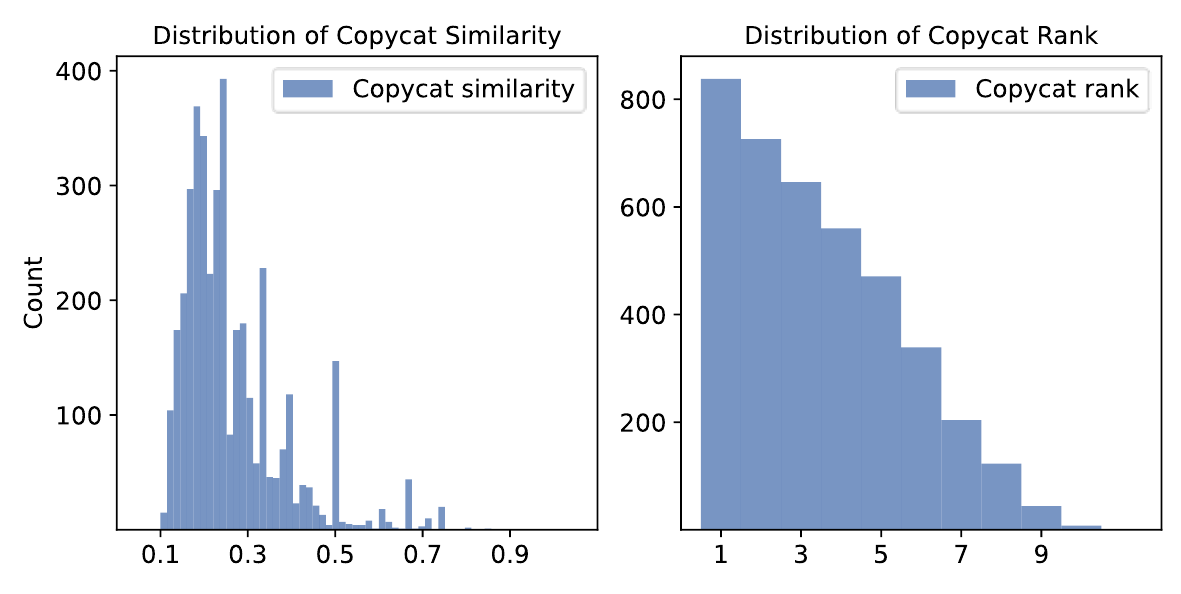}
    \caption{Distribution plots for copycat-related variables}
    \label{fig:distribution_copycat}
\end{figure}

\noindent  \textbf{Statistics for Copycat-related and Franchise Variables.}
Figure \ref{fig:distribution_copycat} shows the distribution of the copycat similarity and the copycat rank variable.  Table \ref{tab:copycat_stats} summarizes the statistics of "copycat" related variables. Of the 35,794 movies, we identify 4,211 copycats, of which 956 are also franchise movies. On average, copycats generate more revenue than non-copycats (Average $\text{log}_{10}$(Revenue) $7.68 > 6.78$ and $6.20 > 5.72$). Franchise movies generate more revenue than non-franchise movies (Average $\text{log}_{10}$(Revenue) $7.68 > 6.20$ and $6.78 > 5.72$). Within the realm of copycats, whether a movie is a franchise or not does not affect its similarity to blockbuster movies much (Similarity $0.25 \approx 0.24$). Furthermore, we find that copycat movies that are part of a franchise tend to have a lower average copycat rank (Rank $0.26 < 0.35$). In other words, their release date is closer to the release date of the corresponding blockbuster. This phenomenon is probably due to the fact that many franchise movies are planned for sequels in the early stages of production, allowing them to secure earlier release dates than their competitors who choose to produce similar movies.


\vspace{-0.3cm}
\begin{table}[h]
\caption{Descriptive statistics of the `copycat-related' variable} \label{tab:copycat_stats}
\resizebox{\columnwidth}{!}{
\begin{tabular}{rrrrr}
\hline
                              & \multicolumn{2}{c}{\# Obs.} & \multicolumn{2}{c}{Avg. $\text{log}_{10}$(Revenue)} \\
                              \cline{2-5} 
                              & \multicolumn{1}{c}{franchise}  & \multicolumn{1}{c}{non-franchise}     & \multicolumn{1}{c}{franchise} & \multicolumn{1}{c}{non-franchise}  \\
                              \cline{2-5} 
\textit{\textbf{Copycat}}     & 956       & 3255       & 7.6853     & 6.2072           \\
\textit{\textbf{Non-Copycat}} & 1,932     & 29,651     & 6.7828     & 5.7287           \\ \hline
\end{tabular}
}
\resizebox{\columnwidth}{!}{
\begin{tabular}{rrrrr}
                              & \multicolumn{2}{c}{Avg. Copycat Similarity} & \multicolumn{2}{c}{Avg. Copycat Rank} \\
                              \cline{2-5} 
                              &\multicolumn{1}{c}{franchise}  & \multicolumn{1}{c}{non-franchise}     & \multicolumn{1}{c}{franchise} & \multicolumn{1}{c}{non-franchise}               \\
                              \cline{2-5} 
\textit{\textbf{\quad \quad Copycat    }}     & 0.2559  & 0.2486   & 0.2653   & 0.3538     \\ \hline
\end{tabular}
}
\end{table}
\vspace{-0.3cm}

\noindent 
\vspace{-0.5cm}
\begin{figure}[h]
    \centering
    \includegraphics[scale = 0.58]{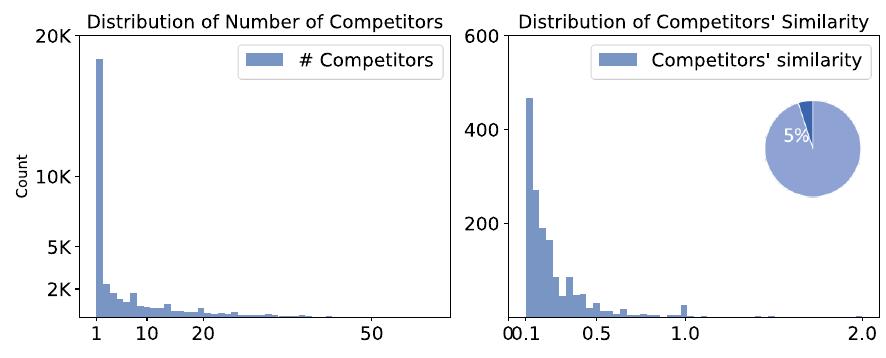}
    \caption{Distribution plots for Competition variables. 
Pie charts represent the proportion of non-zero values in the overall training and validation data. }
    \label{fig:competitor}
\end{figure}
\noindent
\begin{figure*}[t]
    \centering
    \includegraphics[scale = 0.45]{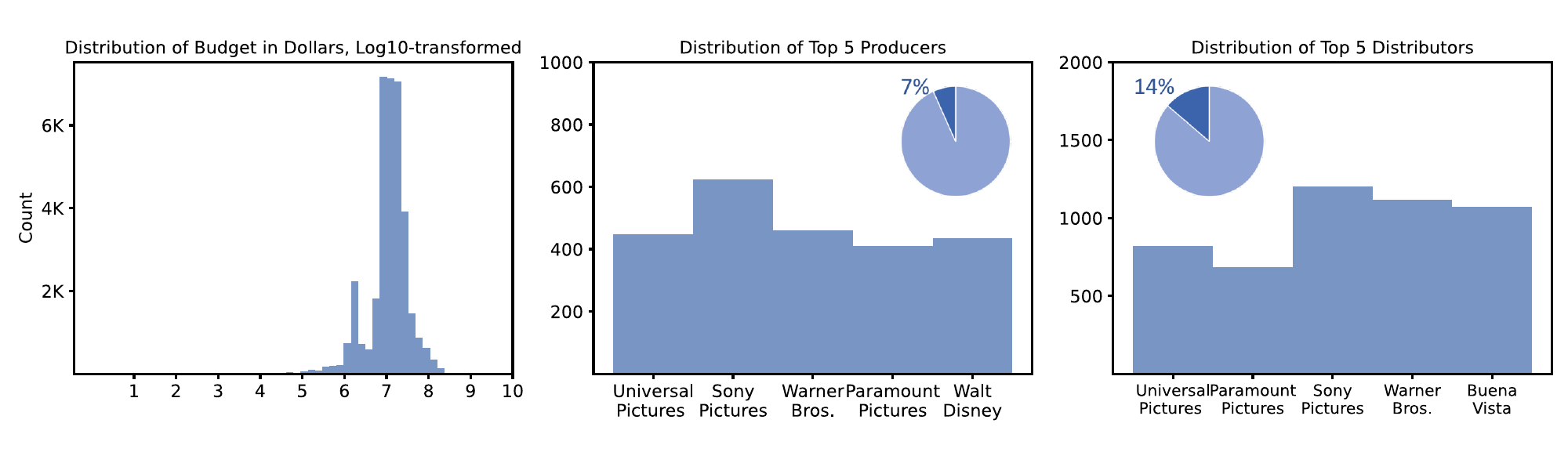}
    \caption{Distribution plots for Investment \& Marketing variables. 
Pie charts represent the proportion of five big names in the overall training and validation data.  }
    \label{fig:investment_marketing}
\end{figure*}

\begin{figure*}[t]
    \centering
    \includegraphics[scale = 0.45]{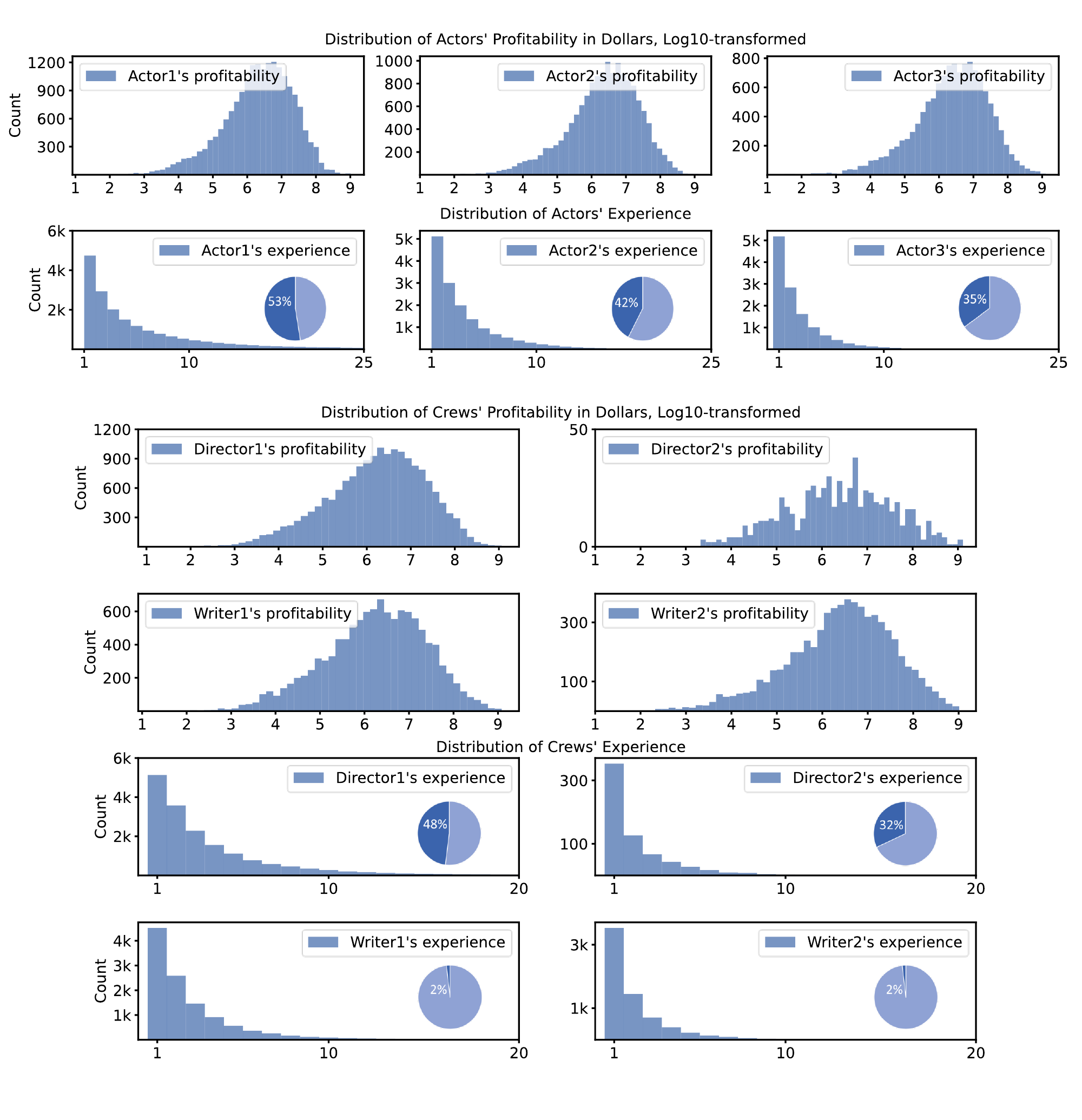}
    \caption{Distribution plots for Star Power variables. Pie charts represent the proportion of non-zero values in the overall training and validation data.  }
    \label{fig:star_power}
\end{figure*}
\noindent

\noindent  \textbf{Competition \& Seasonality.} To capture the effects of changing consumer tastes and holiday seasons, we include the year and month of the movie release as discrete tokens. Further, we model the competition intensity during the release window. We first identify competitors as those released two weeks before and after the current movie and have the same genre. We then sum up the overlap of content keywords, computed using the Jaccard similarity between the current movie and every competitor. Figure \ref{fig:competitor} 
displays the distribution of the number of competitors during the same period, as well as the distribution of their similarity. The pie chart illustrates the proportion of cases with competitors, which is relatively small, indicating that the majority of movies do not face a direct competition from similar movies during the same period.

\noindent \textbf{Investment \& Marketing.} The production budget is often an indicator of the quality of the movie. Here we take the logarithm with a base of 10. We also include the distributor and producer information as tokens, as distributors with greater market power may release movies on more screens and producers may involve varying level of marketing efforts, which increase revenue. Figure \ref{fig:investment_marketing} 
illustrates the distribution of budgets. We have listed the top 5 movie production and distribution companies for the Producer and Distributor categories while grouping the remaining companies into one category.
Familiar names include Warner Bros., Disney, and others. In \S \ref{sec:explanatory} in the main text, we have highlighted the top five production companies (and the distribution companies they own) to investigate their influence.

\noindent  \textbf{Star Power.} We include up to two directors, two writers, and three leading actors in our model. Each person is a unique token whose embedding is trained from scratch. To estimate star power, we calculate the profitability of each person, which is defined as the average of the revenues of all previous movies in which this person has participated in as one of the leading roles. We also estimate the experience of stars by counting the previous movies that this person has participated in. If the star has never been found in any movies as a leading role, these two values would be zero. 
Moreover, we incorporate the gender and age of each actor at the time of the movie release. The same applies to the Director and Writer categories. Figure\ref{fig:star_power}
illustrates the distribution of profitability and experience for cast and crew. 


We show a summary of features in Table \ref{tab:var_desc}. 

\begin{table*}[!h]
 \captionsetup{justification=centering}
\caption{Variable Description}
\label{tab:var_desc}
\centering
\resizebox{0.9\linewidth}{!}{
\begin{tabular}{llll}
\toprule 
 & Description & \makecell[c]{In \\ OLS/RF} &\makecell[c]{In \\ Transformer}  \\\cline{2-4}
\\
\rowgroup{\textbf{Movie Content Related features}}  \\
[0.6ex]
Keywords & \makecell[l]{Keywords, which can be clustered into 1414 clusters} & - & in tokens \\
[0.6ex]
Franchise & The franchise status of the movie & binary & in tokens \\
[0.6ex]
Franchise Name & The franchise collection the movie belongs to & - & in tokens \\
[0.6ex]
Copycat & Movies defined as Copycat or not & binary & in tokens \\
[0.6ex]
Copycat Rank & Copycat movies sorted in chronological order & numerical  & numerical  \\
[0.6ex]
Copycat Similarity & The similarity between Copycat and the original movie & numerical  & numerical  \\
 &  &  &  \\
\rowgroup{\textbf{Genre and MPAA}}  \\
[0.6ex]
Types of Genres & Nineteen different genres & categorical & in tokens \\

MPAA Ratings & \makecell[l]{G, PG, PG-13, R, NC17, NotRated, and N.A. } & categorical & in tokens \\
&  &  &  \\ 
\rowgroup{\textbf{Investment \& Marketing features}} \\
[0.6ex]
Name of Producers & Names of 594 production companies & - & in tokens \\
[0.6ex]
Name of Distributors & Names of 407 distributor companies  & - & in tokens \\
[0.6ex]
Big Producers & \makecell[l]{All the producers grouped into 6 categories, \\ including 5 big companies and the others category} & categorical & - \\
[2ex]
Big Distributors  & \makecell[l]{All the distributors grouped into 6 categories, \\ including 5 big companies and the others category} & categorical & - \\
[2ex]
Budget &  N.A. values replaced by the average value within the same genre & numerical  & numerical  \\
 &  &  &  \\

\rowgroup{\textbf{Star Power features}} \\
Directors Experience & \makecell[l]{The count of  all previous movies that this person \\ has participated in} & numerical  & numerical  \\
[2ex]
Writers Experience & Same as above  & numerical  & numerical  \\
[0.6ex]
Actors Experience & Same as above & numerical  & numerical  \\
[0.6ex]

Directors Profitability &  \makecell[l]{The average of the revenues of all previous movies \\ that this person has participated in} & numerical  & numerical  \\
[2ex]
Writers Profitability & Same as above & numerical  & numerical  \\
[0.6ex]
Actors Profitability & Same as above & numerical  & numerical  \\
[0.6ex]
Director Names & Names of up to two directors  & - & in tokens \\
[0.6ex]
Writer Names & \makecell[l]{Names of up to two writers \\ There are in total 15333 names for directors and writers} & - & in tokens \\
[2ex]
Actor Names & \makecell[l]{Names of up to three actors\\ There are in total 20366 names for actors}  & - & in tokens\\
[2ex]
Actor Ages & Age of each actor  & - & in tokens \\
[0.6ex]
Actor Genders & Gender of each actor  & - & in tokens \\
 &  &  &  \\
\rowgroup{\textbf{Competition \& Seasonality features}} \\
[0.6ex]
Number of Competitors & \makecell[l]{The number of movies in the same genre released during \\ the same period} & numerical  & numerical  \\
[2ex]
Competitors Similarity &The sum of the keywords similarity over Competitors & numerical  & numerical  \\ 
[0.6ex]
Release Year & The year of the movie's first release  & numerical & in tokens \\
[0.6ex]
Release Month & Eleven different months & categorical & in tokens \\
[0.6ex]
\hline
\end{tabular}
}
\end{table*}
\clearpage

\section{More Experimental results}\label{appendix:adjustlossweights}
\vspace{-3cm}

\begin{figure}[htbp]
   \centering
   \includegraphics[scale = 0.5]{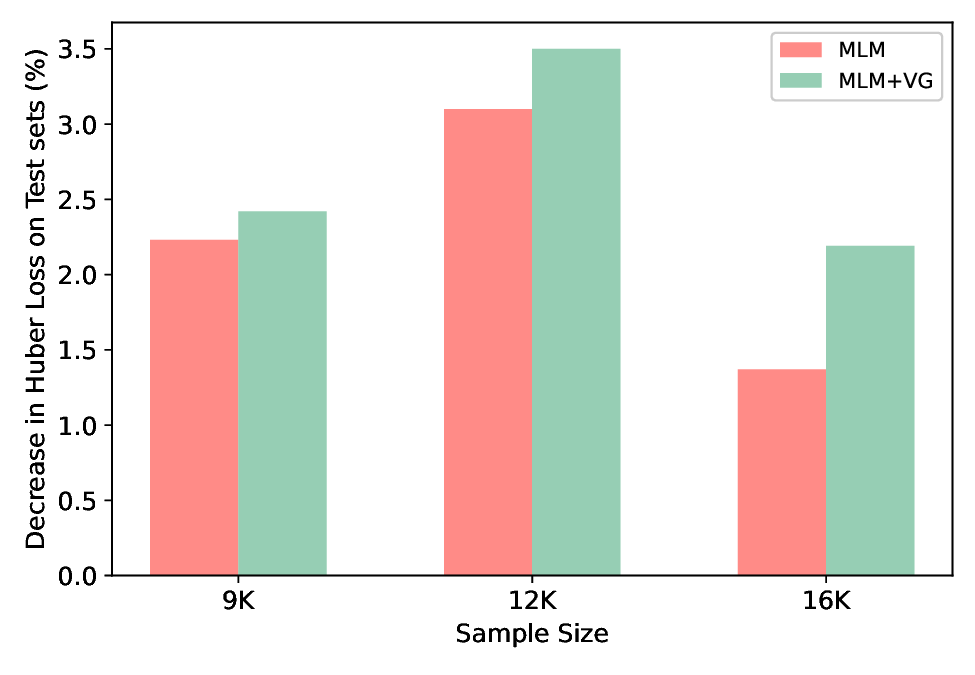}
   \caption{Huber loss decrease between training on randomly sampled data, for both MLM pretraining model and MLM+VG pertaining model with sample size vary. The gap between MLM and MLM+VG grows as the sample size increases.}
   \label{fig:figure_for_report_appendix_tb5}
\end{figure}

\noindent\textbf{Detailed Comparison between MLM and MLM+VG.} In Table \ref{tab:appendix_tb5}, we show detailed numbers comparing the two pretraining objectives, MLM and MLM+VG. For each model, we run three random trials and then compute the average and standard deviation. In Figure \ref{fig:figure_for_report_appendix_tb5}, 
we show relative test loss improvement when the training data size increases. The green bars indicate the average improvements of the MLM+VG model and the red bars indicate the average improvements of the MLM model. We note that as the sample size increases, the gap between the MLM+VG model and the MLM-only models increases, suggesting that the effectiveness of visual grounding increases with data.

\noindent
\vspace{-3cm}
\begin{table}[htbp]
\centering
\small
\caption{ For each model, we run three random trials and then compute the average and standard deviation}\label{tab:appendix_tb5}
\begin{tabular}{cccc} 
\toprule
& & \multicolumn{2}{c}{Average Test Huber Loss\_{(std.)}} \\
\cline{3-4}
& sample size & \makecell[c]{MLM \\ pretraining}  & \makecell[c]{MLM+VG \\ pretraining} \\
\midrule
random samples & 25k         & $0.3097_{(0.0004)}$       & $0.3044_{(0.0005)}$       \\ 
               & 20k         & $0.3180_{(0.0028)}$       & $0.3116_{(0.0007)}$        \\
               & 16k         & $0.3213_{(0.0010)}$       & $0.3201_{(0.0018)}$       \\
               & 12k         & $0.3322_{(0.0002)}$       & $0.3290_{(0.0012)}$       \\
               & 9k          & $0.3415_{(0.0025)}$       & $0.3392_{(0.0015)}$       \\ \midrule
with keywords  & 16k         & $0.3169_{(0.0012)}$       & $0.3131_{(0.0009)}$        \\
               & 12k         & $0.3219_{(0.0007)}$       & $0.3175_{(0.0011)}$         \\
               & 9k          & $0.3339_{(0.0020)}$       & $0.3310_{(0.0012)}$         \\ \midrule
w/o keywords   & 9k          & $0.3498_{(0.0014)}$       & $0.3482_{(0.0002)}$      \\ \bottomrule
\end{tabular}
\end{table}
\vspace{-2cm}


\vspace{5cm}

\noindent\textbf{Poster Retrieval Results.} In the main text, we mentioned that our method results in a better keyword representation than the BERT pertaining embedding. We can check the meaning of our embedding by using it to retrieve the most similar posters. 
We are displaying some poster retrieval results here that could not be included in the main text due to the page limit. These results demonstrate that the keyword embeddings we obtained through pertaining and self-supervised learning are meaningful.

\begin{figure}[h]
    \centering
    \includegraphics[scale = 0.35]{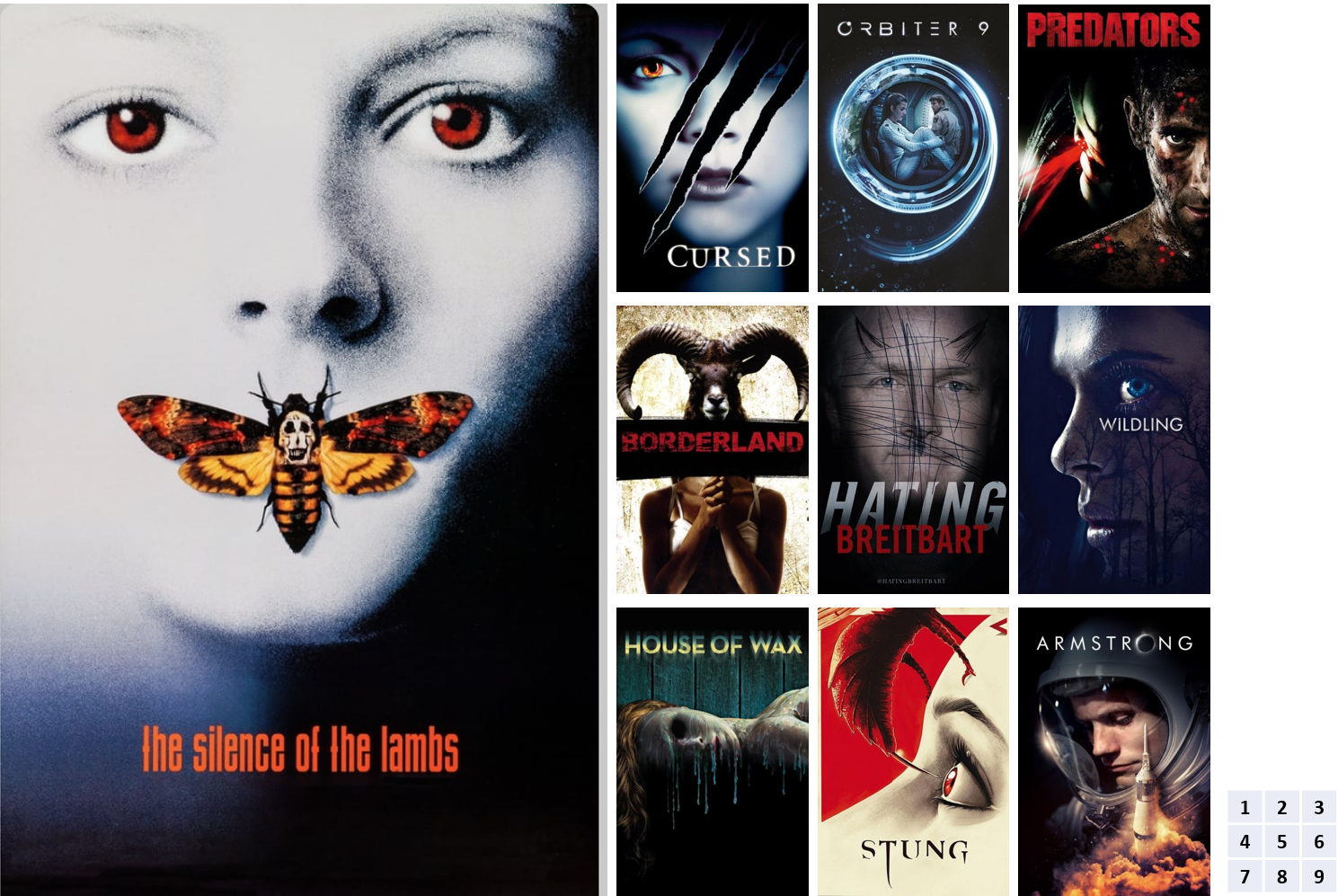}
    \caption{Poster retrieval (top 9) using the keyword `psycho' and a typical Thriller movie \emph{The Silence of the Limb (1991)}}
\end{figure}
\vspace{-0.3cm}
\begin{figure}[h]
    \centering
    \includegraphics[scale = 0.35]{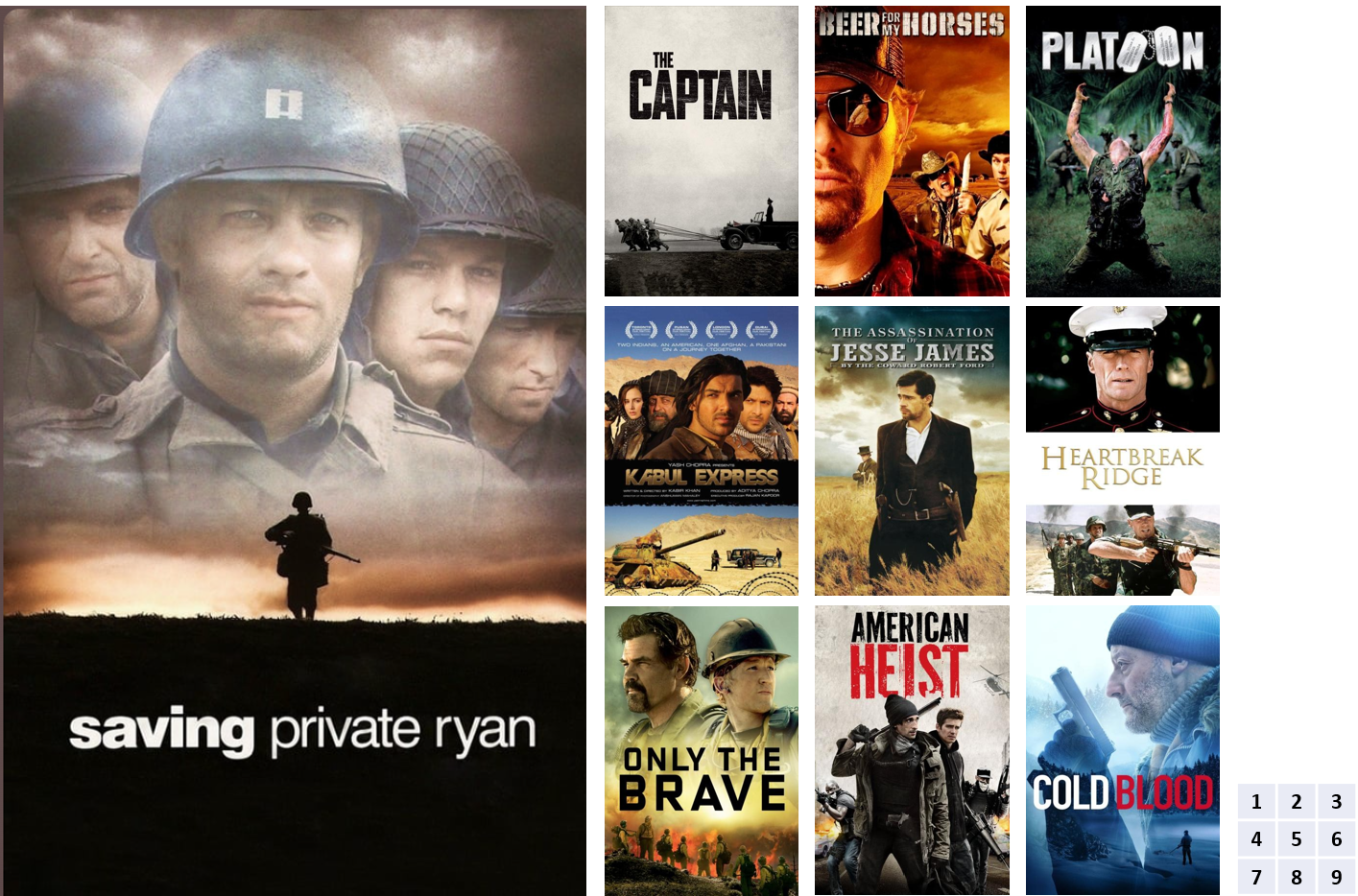}
    \caption{Poster retrieval using the keyword `war' and a typical War movie \emph{Saving Private Ryan (1998)}}
\end{figure}
\vspace{-0.3cm}
\begin{figure}[h]
    \centering
    \includegraphics[scale = 0.35]{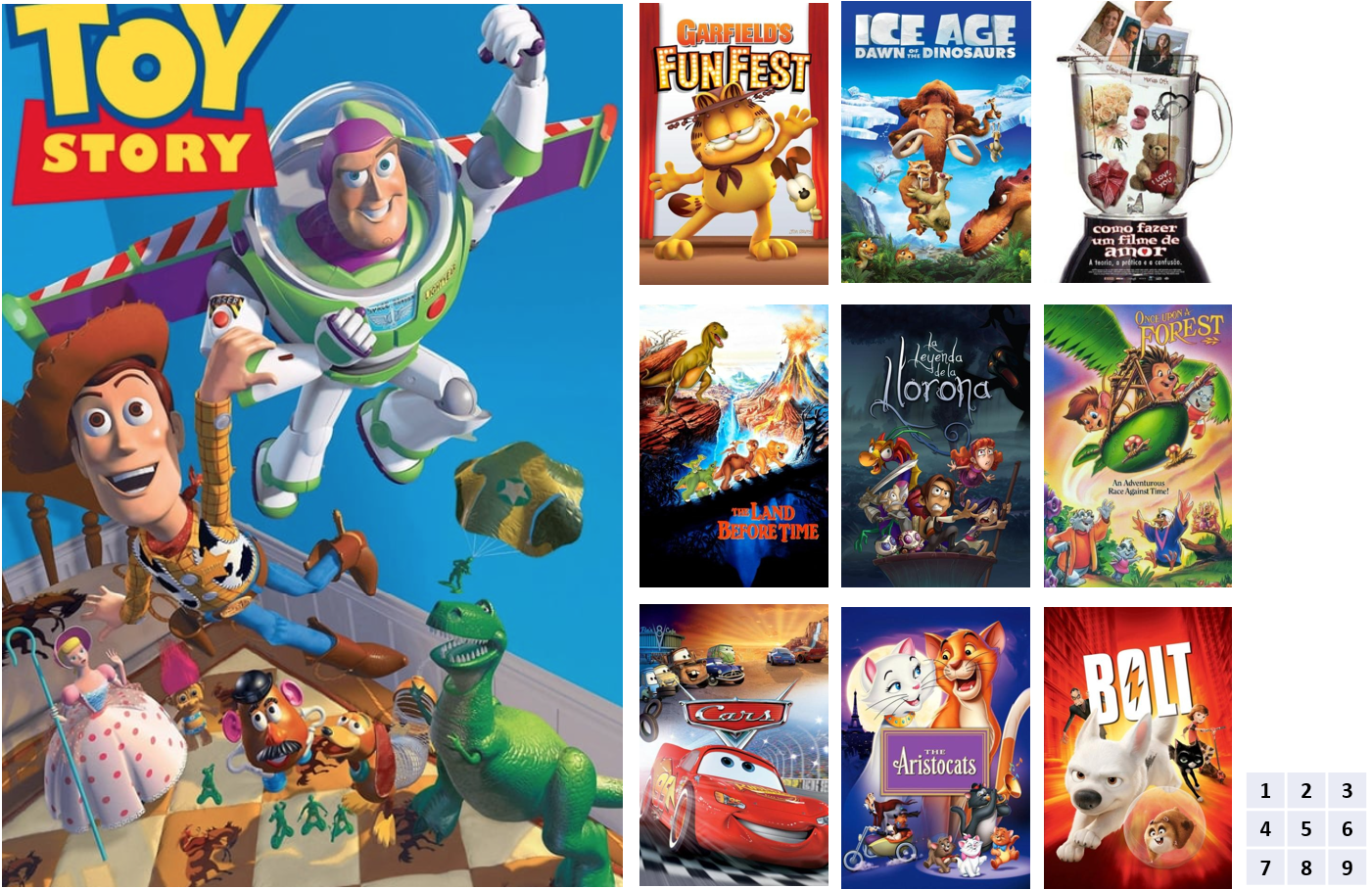}
    \caption{Poster retrieval using the keyword `friendship' and a typical Family \& Animation movie \emph{Toy Story (1995)}}
\end{figure}
\noindent
\end{appendices}

\end{document}